\documentclass[letterpaper, 10 pt, conference]{ieeeconf}  

\IEEEoverridecommandlockouts                              
\usepackage{graphicx,subfigure}
\usepackage{color}
\usepackage{CJK}
\usepackage{framed}
\usepackage{ulem}
\usepackage{cite}
\usepackage{caption}
\usepackage{amsmath}
\usepackage{amssymb}
\usepackage{commath}
\usepackage[ruled,norelsize,vlined,linesnumbered]{algorithm2e}
\makeatletter
\newcommand{\removelatexerror}{\let\@latex@error\@gobble}
\makeatother
\graphicspath{ {Figures/} }

\title{\LARGE \bf
Reconfigurable Multi-UAV Formation Using Angle-Encoded PSO
}

\author{V.T. Hoang, M.D. Phung, T.H. Dinh, Q. Zhu, Q.P. Ha 
\thanks{The authors are with School of Electrical and Data Engineering, Faculty of Engineering and Information Technology (FEIT), University of Technology Sydney (UTS), 81 Broadway, Ultimo NSW 2007, Australia
        {\tt\small \{VanTruong.Hoang, ManhDuong.Phung, TranHiep.Dinh, Qiuchen.Zhu, Quang.Ha\}@uts.edu.au}}%
}

\begin{document}
\bibliographystyle{IEEEtran}

\maketitle
\thispagestyle{empty}
\pagestyle{empty}

\begin{abstract}

In this paper, we propose an algorithm for the formation of multiple UAVs used in vision-based inspection of infrastructure. A path planning algorithm is first developed by using a variant of the particle swarm optimisation, named $\theta$-PSO, to generate a feasible path for the overall formation configuration taken into account the constraints for visual inspection. Here, we introduced a cost function that includes various constraints on flight safety and visual inspection. A reconfigurable topology is then added based on the use of intermediate waypoints to allow the formation to avoid collision with obstacles during operation. The planned path and formation are then combined to derive the trajectory and velocity profiles for each UAV. Experiments have been conducted for the task of inspecting a light rail bridge. The results confirmed the validity and effectiveness of the proposed algorithm.


\end{abstract}

\section{INTRODUCTION}

The use of Unmanned Aerial Vehicles (UAVs) for civil infrastructure monitoring and inspection is receiving increasing interest recently due to its various benefits over traditional inspection methods such as risk and budget reduction, traffic free and flexible operation. In practice, a number of UAV systems have been successfully deployed to periodically monitor the condition of bridges, buildings, wind turbines, gas pipelines, etc. \cite{rathinam2008vision}. However, the need to inspect large structures with high accuracy and small completion time are posing new challenges that transcend the use of a single UAV to a formation-based approach. The approach features multiple UAVs coordinating in a certain configuration with capabilities to adapt itself to the operating environment.   

In the literature, approaches for formation problems can be categorized into the consensus-based \cite{Yan2017,dong2015, he2018}, artificial potential function-based \cite{kim2006, do2007, Liao2017}, leader-follower (L-F) \cite{semsar2008, panagou2014, yu2016}, observer-based \cite{Nguyen2004}, behavior-based and virtual structure \cite{sharma2009, wang2013} methods. Among them, the virtual structure formation is more relevant for visual inspection as it enables quick reconfiguration which is essential in cases of failure in communication, sensor/actuator, flight path constraints or even the loss of one or more agents. The general idea of the reconfigurable problem includes defining a set of new separation distance, position, and other parameters that are suitable for specific mission requirements, and a relevant process to achieve that configuration \cite{wang1998, wang1999}.  Optimal algorithms were mostly preferable to solve reconfiguration, such as semianalytic approach \cite{sauter2012}, PSO \cite{duan2010, sui2017} or hybrid PSO \cite{duan2013, wang2013}, nonlinear programming \cite{ma2010} and hierarchical evolutionary \cite{wang2012} trajectory planning. Other methods also contribute their advantages to solve the formation reconfiguration problem, including obstacle and collision avoidance, i.e., potential field \cite{paul2008, feng2018}, distributed controller with a failure detection logic \cite{seo2012} and interior point algorithm \cite{jian2016}. For visual inspection, the reconfigurable problem, however, involves new constraints on data collection which require further investigation such as the constraints on waypoints for photos taken, the distance between UAVs and inspecting surfaces, or the overlap among photos for stitching. 

In this paper, we propose a new algorithm for reconfigurable formations based on the angle-encoded PSO. We begin with the use of a 3D representation of the surface to be inspected and its neighboring environment to determine a set of intermediate waypoints (WPs) necessary for the UAVs to travel through in order to collect sufficient data for inspection. A formation shape is then defined at these WPs so that multiple UAVs are feasible to operate. A set of new constraints is additionally proposed to increase collision avoidance ability and task performance. Based on that, an optimal path for the centroid of the formation is produced by employing the $\theta$-PSO path planning algorithm \cite{hoang2018iros}. Finally, trajectories for individual UAVs will be achieved by integrating the generated path with the selected reconfiguration shapes. Experiments have been carried out with the results confirmed the validity and effectiveness of the proposed approach.


The main contribution of this work is an augmentation of the proposed optimisation algorithm for path planning \cite{hoang2018iros} with additional constraints based on intermediate waypoints to satisfy the requirements for safety and task efficiency given a number of formation types that can be reconfigurable during operations. The new formation path planning technique can therefore (i) generate safer flights of the UAV group, and (ii) improve the formation flexibility as well as its capacity for inspection tasks.

The paper is organised as follows. Section \ref{path} presents the design of the path planning algorithm using $\theta$-PSO. The proposed reconfigurable configurations and implementation of trajectory planning for UAVs are described in Section \ref{formation}. Experimental results are introduced in Section \ref{results}. The paper ends with conclusions and recommendations for further study described in Section \ref{conclusion}.

\section{PATH PLANNING FOR VISUAL INSPECTION}  \label{path}
Our approach begins with the generation of a flyable path for the triangular centroid of a multiple-UAV formation conducting surface inspection tasks. This path will be further adjusted to include the reconfiguration for the formation as presenting in next sections. As we aim to create an optimal path with quick time convergence, the angle-coded PSO ($\theta$-PSO) is used \cite{hoang2018iros}. In $\theta$-PSO, the location of particles is encoded by the angle of the formation centroid and their movements are described as:
\begin{align}\label{EqThePSO}
\begin{cases} \Delta \theta_{ij}^{k+1} = w \Delta \theta_{ij}^k + c_1 r_{1i}^k (\lambda_{ij}^k - \theta_{ij}^k) + c_2 r_{1i}^k (\lambda_{gj}^k - \theta_{ij}^k) \\
\theta_{ij}^{k+1} = \theta_{ij}^k + \Delta \theta_{ij}^{k+1}, (i=1,2,...,N; j=1,2,...,S)\\
x_{ij}^k = \dfrac{1}{2}\left[(x_{max} -x_{min}) sin \left( \theta_{ij}^k \right) + x_{max} + x_{min} \right], \end{cases}
\end{align}
where subscript $k$ represents the iteration index; $w$, $N$ and $S$ are the inertial weight, total number of particles and the dimension of the search-space, respectively; $r_1$ and $r_2$ are pseudo-random scalars; $c_1$ and $c_2$ are the gain coefficients; $\theta_{ij} \in [-\pi/2, \pi/2]$  represents the phase angle, and $\Delta \theta_{ij} \in [-\pi/2, \pi/2]$ is the increment of $\theta_{ij}$ in the $j$th dimension of the particle $i$; $\lambda_g = [\lambda_{g1}, \lambda_{g2},\ldots, \lambda_{gS}]$ and $\lambda_i = [\lambda_{i1}, \lambda_{i2}, \ldots, \lambda_{iS}]$ represent the global and personal best positions, respectively; $x_{ij} = f(\theta_{ij})$ is the $j$th dimension of the $i$th particle's position; $x_{max}$ and $x_{min}$ are the upper and lower constraints of the searching space. 

For the $\theta$-PSO to find the global optimum solution corresponding to the shortest flyable path, it is essential to define a proper cost function incorporating a number of constraints relating to the maneuverability of UAVs, operating space, task performance, and collision avoidance. In this work, the objective function is defined in the following form:
\begin{equation} \label{path_cost}
J_F(T_{Fi}) = \sum \limits_{m=1}^3 \beta_m J_m(T_{Fi}),
\end{equation}
where $T_{Fi}$ is the $i$th path for the formation to be judged; $\beta_m$ represents the weighting factor selected for the corresponding cost component; and $J_m(T_{Fi})$, $m=1..3$, are respectively the cost components correlated with the length of a path, obstacle avoidance, and operating height. 

To determine $J_m(T_{Fi})$, the centroid path $T_{Fi}$ is divided into $L_i$ segments, where $L_i$ is chosen to be sufficiently large so that each segment can be considered to be straight and represented by coordinates of ending nodes $P_{i,l} = \{x_{i,l},y_{i,l},z_{i,l}\}, l= 0 .. L_i$. The path length cost component $J_1$ is then computed for all segments as:
\begin{align}
J_1(T_{Fi}) &= \sum \limits_{l=1}^{L_i} \norm{{P_{i,l} - P_{i,l-1}}},
\end{align}
where $\norm{.}$ denotes the Euclidean norm.

To form $J_2$ for collision avoidance, let $K$ be the number of all obstacles within the operation space. The overall violation cost scaled across all path segments and $K$ obstacles then can be obtained as:
\begin{equation}\label{J_2}
J_2(T_{Fi}) =  \frac{1}{{L_i}K} \sum \limits_{l=1}^{L_i}  \sum \limits_{k=1}^K \text{max} (1-\dfrac{d_{l,k}}{r^S_{l,k}},0),
\end{equation}
where $d_{l,k}$ is the actual distance between the $k$th obstacle and the midpoint of segment $l$, and $r^S_{l,k}$ is a safe radius of the formation w.r.t. the obstacle $k$.

Finally, the cost $J_3$ relating to the altitude constraints that restrict the UAVs to travel within a predefined height range, represented by the minimum and maximum values, $z_{\text{min}}$ and $z_{\text{max}}$ can be expressed as:
\begin{align}
\begin{cases}
&J_3(T_{Fi}) =  \sum \limits_{l=1}^{L_i} \delta_{l}\\
&\delta_{l} = \begin{cases}
			z^M_{l} - z_{\text{max}},  &\text{if } z^M_{l} > z_{\text{max}}\\
			0,  &\text{if } z_{\text{min}} \leq z^M_{l} \leq z_{\text{max}}\\
			z_{\text{min}} - z^M_{l},  &\text{if } 0< z^M_{l} < z_{\text{min}}\\
			\infty,  &\text{if } z^M_{l} \leq 0.
			\end{cases}
\end{cases},
\end{align}
where $z^M_l$ represents the height of segment $l$.

\section{RECONFIGURABLE FORMATION}  \label{formation}
In this section, the path planning result is updated with a reconfigurable strategy in order to complete a safe trajectory for an individual UAV during its operation in formation. It begins with a triangle formation model with some potential transforming shapes, followed by the fix waypoint selection and ends with the individual trajectory generation for each UAV and the overall algorithm.
\subsection{Introduction of UAV Formation Topologies}  \label{form_model}
\begin{figure}
	\centering
	\includegraphics[width=7cm]{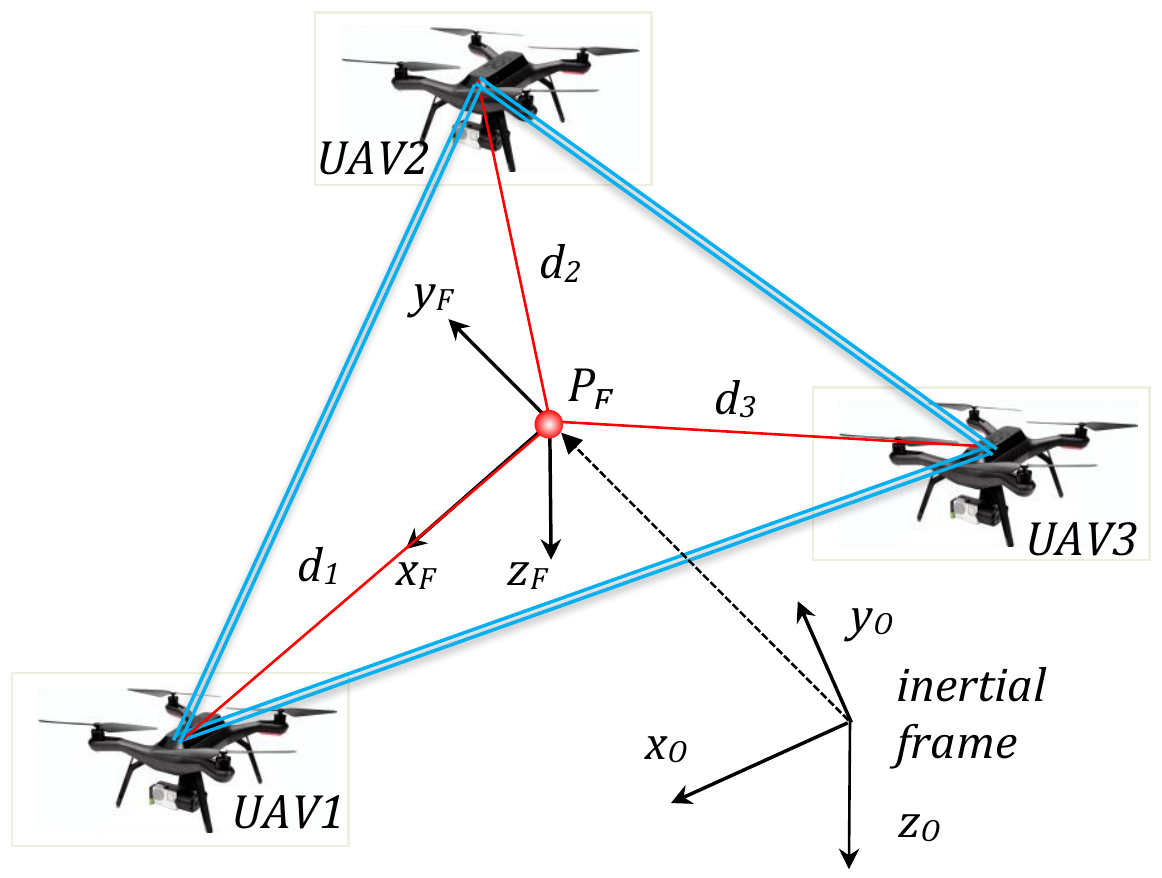}
	\caption{UAV formation frames}
	\label{form_fig}
\end{figure}

Figure \ref{form_fig} illustrates the two frames that represent a triangular formation created by three UAVs, the formation and the inertial frames. The formation frame, $\{x_F, y_F, z_F\}$, is determined such that its origin $P_F$ is selected to be coincident with the triangle centroid. Denoting $P_n = \{x_n, y_n,z_n\}$, $n=1,2,3$, as the position of UAV$_n$ and $d_n$ as the distance between UAV$_n$ and $P_F$. The formation is then represented by the position $P_F$ computed as:
\begin{equation} \label{form_model}
P_F = \dfrac{1}{3} \sum\limits_{i=1}^3 P_n,
\end{equation}
and the radius is given by $r_F = \text{max}(d_{n})$, while the rotation matrix, $R_{IF}$, relating the formation and inertial frames is given in \cite{hoang2018iros}.

\begin{figure}[!htbp]
	\centering
	\includegraphics[width=8.3cm]{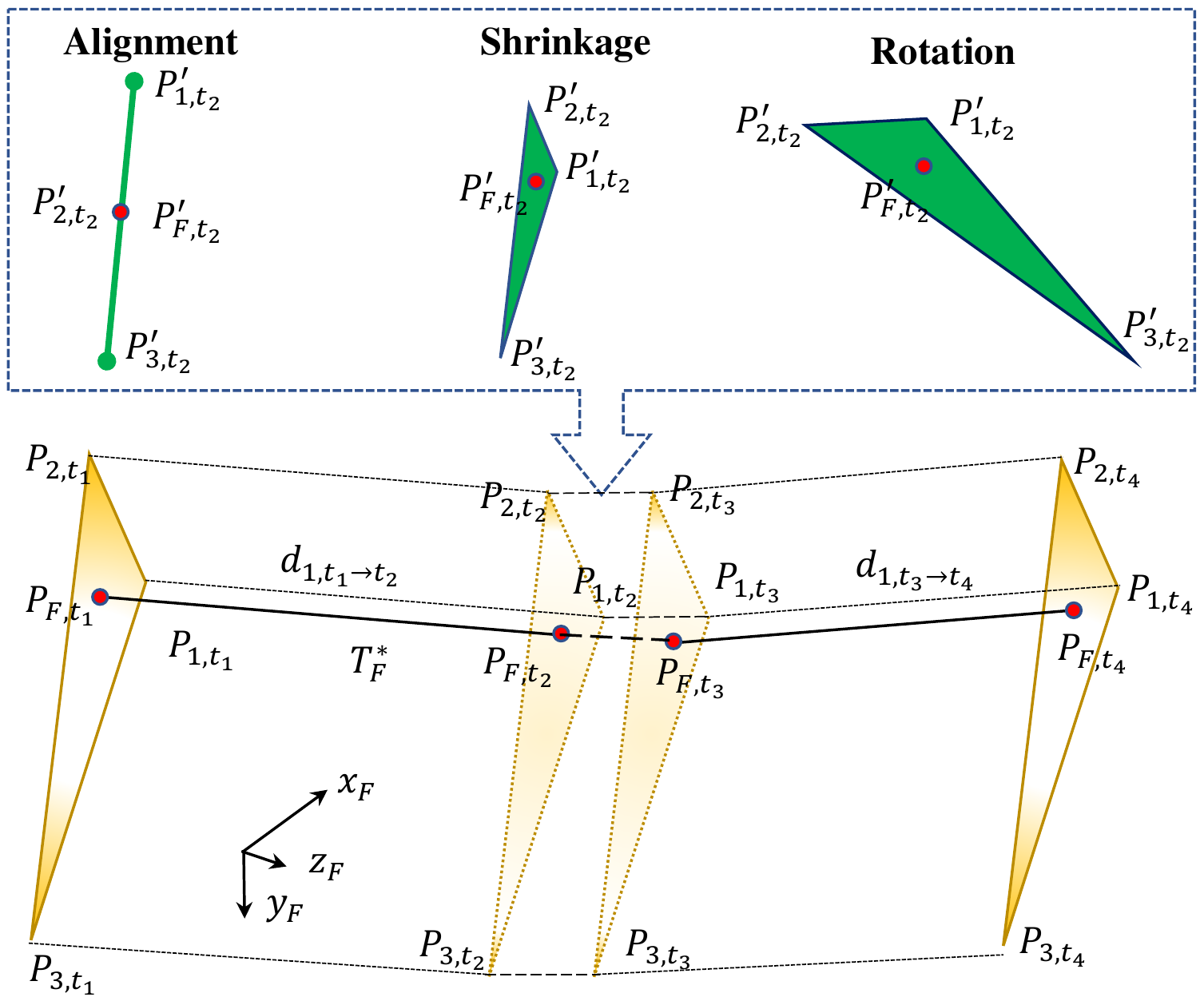}
	\caption{Reconfigurable formation}
	\label{trans_all}
\end{figure}

The triangular formation can be used to coordinate the UAVs for inspection tasks given that the formation is managed as a rigid body. Under that assumption, the path generated in Section \ref{path} can be directly used as the reference for the formation centroid. In practice, the rigid-body assumption is not always held as the UAVs may need to change the formation shape to adapt to the operating environment. For example, a narrow passage or an unwanted obstacle may require the UAVs to fly in a row or column instead of the triangular. As explained in Fig. \ref{trans_all} the following reconfigurations are considered in this study:
\begin{itemize} \label{formlist}
	\item[$\cdot$] Alignment: The UAVs form a line. It is used for the scenarios of appearing narrow passages/obstacles that is only possible for a single UAV to pass.
	\item[$\cdot$] Rotation: The UAVs rotates as a rigid body structure to preserve the formation shape. It allows the UAVs to quickly turn back to the previous formation configuration.
	\item[$\cdot$] Shrinkage: The UAVs fly toward the formation centroid while maintaining the formation shape. This configuration is used in case of required to maintain the overlap among photos taken. 
\end{itemize}

\subsection{Reconfiguration with Intermediate Waypoints}
In order to reconfigure, the UAVs need to re-route their flying paths through adjacent space and thus require intermediate waypoints (IWPs). To identify those waypoints, additional constraints are required as follows:
\begin{itemize}
\item[1.]  The distance between UAVs must be within the communication range but not smaller than two times of the UAV radius: 
\begin{equation} \label{eq_com_sep}
d_{\text{com}} \geq d(P_m, P_n) \geq 2 r_Q,
\end{equation}
where $d_{\text{com}}$ is the communication range, and $r_Q$ is the safe radius of a UAV, and $d(P_m, P_n)$ is the distance between UAV$_m$ and UAV$_n$.
\item[2.] The UAVs must fly within a certain distance to the inspecting surface:
\begin{equation} \label{eq_inter_safe}
d_n^s \in [d_{\text{min}}^s, d_{\text{max}}^s], \hspace{0.2cm} n \in \{1,2,3\},
\end{equation}
where $d_n^s$ is the distance from UAV$_n$ to the inspecting surface, $d_{\text{min}}^s$ and $d_{\text{max}}^s$ are respectively the minimum and maximum distances from a UAV to the surface.
\end{itemize}

Assume that each obstacle in the working environment of UAVs is modelled as a cylinder with the center's coordinate $C_k$, radius $r_k$ and height $z_k$. For any different obstacles $p$ and $q, \forall p \neq q, \hspace{0.1cm} p,q \in \{1, ..., K\}$, we have their radii $r_p$, $r_q$, and centre coordinates $C_p (x_p, y_p)$, $C_q (x_q, y_q)$, respectively. Denoting $P_p$ and $P_q$ as intersection points between the straight line created by $C_p$ and $C_q$ and the two circles $(C_p, r_p)$ and $(C_q, r_q)$, we determine location of the $j$th IWP as the midpoint of $P_pP_q$:
\begin{equation}\label{eqFWP}
C_j =\begin{cases} \dfrac{1}{2}(P_p+P_q), \hspace{0.1cm}  &\text{if } r^S_n \leq d_{p,q} < r^S_{l,k}\\
\varnothing &\text{otherwise}
\end{cases},
\end{equation}
where $C_j = (x_j, y_j)$ is the coordinates of the $j$th IWP in the horizontal plane, $d_{p,q}$ is the smallest distance between the two adjacent obstacles, and $r^S_n$ is the safe radius of UAV$n$. Finally, the cost function (\ref{path_cost}) need be updated to include the cost caused by the intermediate waypoints as follows:
\begin{equation}\label{cost_function}
J(T_{Fi}) = J_F(T_{Fi}) + J_R(T_{Fi}),
\end{equation}
where $J_R(T_{Fi})$ represents the distance from intermediate waypoints to path segments: 
\begin{equation}\label{J4}
J_R(T_{Fi}) = \frac{1}{{L_i} M_j} \sum \limits_{l=1}^{L_i} \sum \limits_{j=1}^{M_j} \sqrt{(x_l - x_j)^2 + (y_l - y_j)^2},
\end{equation}
where $M_j$ is the total number of IWPs.

At each IWP, the formation is reconfigured by changing positions of the UAVs to a designated position in the newly selected shape. The UAVs then come back to their original defined position after passing those IWPs. Hence, the changing shape can be divided into two phases, transformation and reconfiguration, conducting between time intervals $[t_1,t_2]$ and $[t_3,t_4]$ respectively as shown in Fig.\ref{trans_all}. The new shape is maintained between those phases, from $t_2$ to $t_3$, to keep the UAVs safe while travelling inside the narrow passage. The next step is to find a set of positions, $P_n$ for each UAV$_n$ such that the planned trajectory of the whole formation, represented by the formation centroid $P_F$, and the formation shape are preserved.

\subsection{Reference Trajectory Generation} \label{I_planning}
Given the optimized path, $P^*_F$, produced by the $\theta$-PSO for the centroid of the triangular formation in which the intermediate waypoints for reconfiguration have been included, specific paths for each UAV can be computed based on the formation model presented in (\ref{form_model}). 

For IWP$_j$, a set of new waypoints $P^j_F = [P_{F,t_1},..,P_{F,t_4}]$ for the formation is generated. Depending on the defined position in the reconfiguration shape and $P^j_F$, a set of waypoints for UAV$_n$,  $P^j_n = [P_{n,t_1},..,P_{n,t_4}]$, is also computed. Let $\Delta{P_n}$ be the set of desired difference in position between UAV$_n$ and the formation centroid at time $t$:
\begin{equation}\label{distance}
\Delta{P_n} = (P_n \cup P^j_n) - (P_F\cup P^j_F),
\end{equation}
This difference is calculated in the inertial frame as:
\begin{equation}
\Delta{P^I_n} = R^{-1}_{IF}(t)\Delta{P_n},
\end{equation}
where $R_{IF}$ is the rotation matrix. The flying path for each UAV is then given by:
\begin{equation} \label{re_pos}
P^*_n = P^{*}_{F} + \Delta P_n.
\end{equation}

Finally, by combining the results of the path planning process and the reconfigurable algorithm, the completed set of trajectory commands for the $n$th UAV is determined as:
\begin{equation}\label{traj_com}
T_n = [P^*_n, V_n]^T,
\end{equation}
where $V_n$ is the velocity profile set for UAV$_n$. This command set will be uploaded to the onboard controller of the $n$th UAV for trajectory tracking.

\subsection{Algorithm Implementation}
The implementation of our reconfigurable formation algorithm can be described by the pseudo code in Fig. \ref{figPSOpseudocode}. It starts with the initialisation of the inspection surface, working space, obstacle positions, flight constraints and $\theta$-PSO parameters. The $\theta$-PSO is then executed based on the cost function (\ref{cost_function}) to generate an optimal path for the formation centroid. At each IWP, the chosen formation shape is the basis to compute the new set of positions for UAV$_n$ w.r.t their corresponding positions of the centroid. The distance error, $\Delta d_{n,t_1 \rightarrow t_2}$, is found by comparing between the travel distances, $d_{n,t_1 \rightarrow t_2}$ and $d'_{n,t_1 \rightarrow t_2}$, of the nominal and transformation shape, respectively. The ground velocity increment $\Delta V_{n,t_1 \rightarrow t_2}$ is found based on $\Delta d_{t_1 \rightarrow t_2}$ and the transformation time $t_t$ computed from the planned path. A similar process is applied for the period of $[t_3,t_4]$. 

\begin{figure}[h!]
	{\fontsize{10}{10}\selectfont
	\removelatexerror
		\begin{algorithm}[H]
		\SetAlgoLined
			\tcc{Preparation:}
			\quad Determine the inspection surface(s)\;
			\quad Identify boundaries of the working space\; 
			\quad Identify obstacle set $K$\;
			\quad Group all the above data and save in a common file (init file)\;
			\tcc{Initialisation:}
			\quad Initialise the working environment by loading the init file to global memory\;
			\quad Initialise constraints, i.e., $r_Q, d_{com}, d^s_{min}, d^s_{max}$\; 
			\quad Determine locations of IWPs using Eqs. (\ref{eq_com_sep}), (\ref{eq_inter_safe}), and (\ref{eqFWP})\;
			\quad Initialise $\theta$-PSO parameters\;
			\quad Generate a random path to connect the start and target waypoints\;
			\quad Set $\theta_{ij} \in [-\pi/2, \pi/2]$ and $\Delta \theta_{ij} \in [-\pi/2, \pi/2]$\;
			\tcc{Path Planning:}
			\ForEach {$i < $\textit{(swarm\_iteration)}} {
				\ForEach {$ j < $ \textit{(swarm\_population)}} {
					Compute new value of $\Delta \theta_{ij}$; \tcc*[f]{using 1st equation in (\ref{EqThePSO})}\\ 
					Compute new value of $\theta_{ij}$; \tcc*[f]{using 2nd equation in (\ref{EqThePSO})}\\ 
					Compute new position; \tcc*[f]{using 3rd equation in (\ref{EqThePSO})} \\		
					Check \textit{Violation} cost\;
					Evaluate each path based on the \textit{Best\_Costs} and \textit{Violation} cost\;
					Update each \textit{particle\_personal\_best} and the \textit{global\_best} positions\;
				}
				\quad Update \textit{global\_best} and \textit{Violation} costs\;
			}
			\quad Save \textit{global\_best} and \textit{Violation} cost\;
			\quad The optimized path is achieved when the maximum number of iterations is reached.\\
			\quad Generate the individual path for UAV$_n$.\\
			\tcc{Path generation:}
			\ForEach {UAV$_n$} {			
				\ForEach {$j =1$ to $M_j$} {
				\quad Select a relevant formation shape\;
				\quad Compute the new position set $P^j_n$\;
				\quad Compute the mission time $t_t$ and $t_r$\;
				\quad Determine $\Delta d_{n,t_1 \rightarrow t_2}$ and $\Delta d_{n,t_3 \rightarrow t_4}$\;
				\quad Compute $\Delta V_{n,t_1 \rightarrow t_2}$ and $\Delta V_{n,t_3 \rightarrow t_4}$\;                                            
				\quad $P^*_n \leftarrow P^{*}_{F}, \Delta P_n$ \tcc*[f]{using (\ref{distance})-(\ref{re_pos})}\; 
				\quad $V_n \leftarrow \Delta V_{n,t_1 \rightarrow t_2}, \Delta V_{n,t_3 \rightarrow t_4}$\;
				} 
			\quad $T_n \leftarrow P_n^*$, $V_n$. \tcc*[f]{using (\ref{traj_com})}.			
			} 
		\end{algorithm}
	\caption{Pseudo code for reconfigurable trajectory generation process.}
	\label{figPSOpseudocode}	
  }
\end{figure}
\section{EXPERIMENTS} \label{results}

We have conducted a number of experiments to evaluate the validity and efficiency of the proposed algorithm. The setup and results are presented below.

\subsection{Experimental setup}
The task designated in our experiments is to inspect different surfaces of a light rail bridge employing three identical UAVs. The UAVs used are the 3DR Solo drones retrofitted with inspection cameras and communication boards \cite{hoang2018iros}. The onboard low-level controllers for these drones have been addressed in \cite{Hoang2017AT}. The operation space is chosen in the rectangular area with two opposite corners at GST coordinates of $\{ -33.87601, 151.191182\}$ and $\{-33.875086, 151.192676\}$. Therein, actual obstacles are identified and the mission of UAVs is to inspect the surface represented by their poles numbered from (1) to (12). These obstacles include  a pole (2), a light pole (4), bridge piers (1, 3, 5, 6), power poles (8, 10), and a tree (9).

The initial configuration for inspection was a triangle formation with initial positions of the UAVs relative to the formation centroid to be $\Delta T_1 = [0, 2, 0]$ m, $\Delta T_2 = [-2, -1, 0]$ m and $\Delta T_3 = [2, -1, 0]$ m. The lower and upper bounds of altitudes between the UAVs and the ground are $z_{\text{max}} = 15$ m and $z_{\text{min}} = 7$ m, respectively. The UAVs are required to fly within the relative distance to the inspected structure as $d_{n,s} \in [1,5]$ m. The formation is set to flight at a constant ground velocity of 3 m/s.  For $\theta$-PSO, the swarm size, number of waypoints, and number of iterations are respectively selected as 100, 7, and 150.

\subsection{Results}
\begin{figure}[!htbp]
	\centering
	\includegraphics[width=8.5cm]{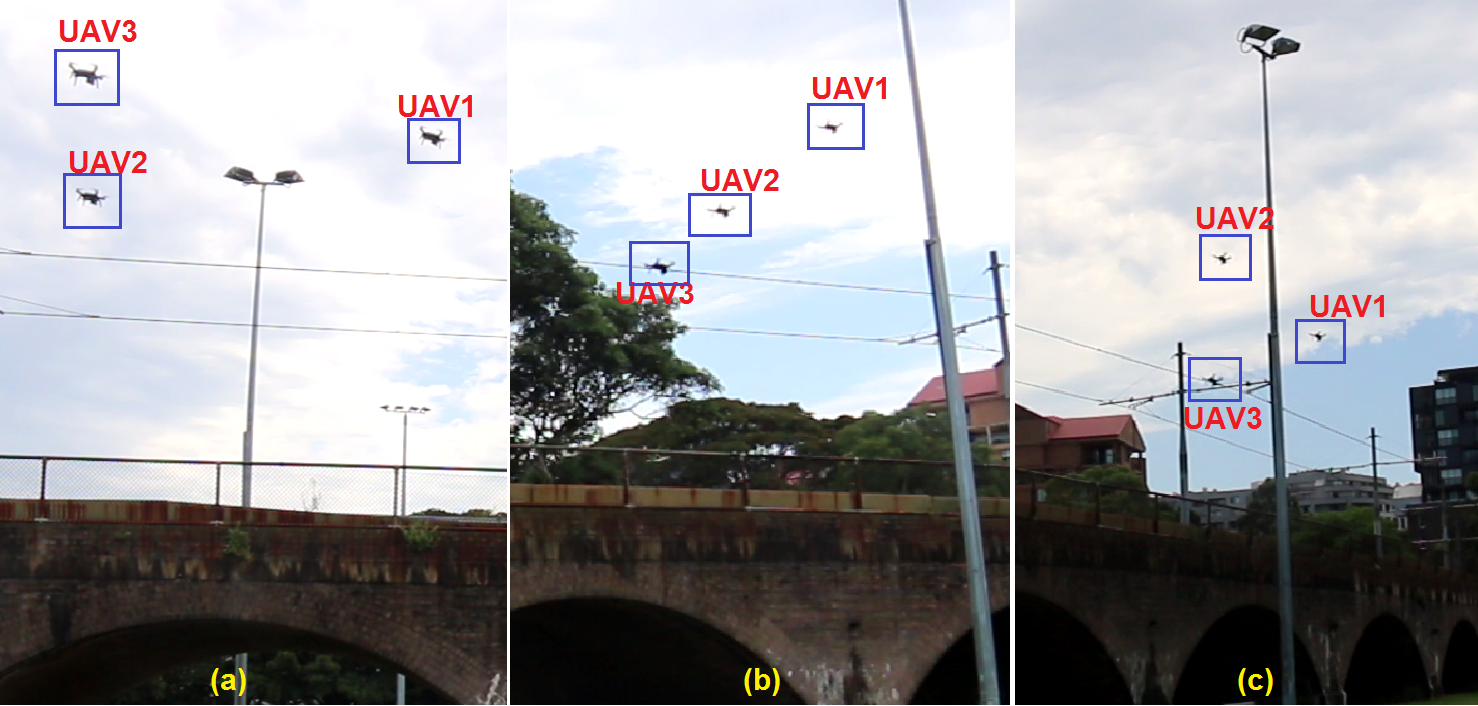}
	\caption{Transformation of triangular formation (a) to alignment (b), and rotation (c) configurations}
	\label{real_form}
\end{figure}

\begin{figure}
	\centering
	\includegraphics[width=8.5cm]{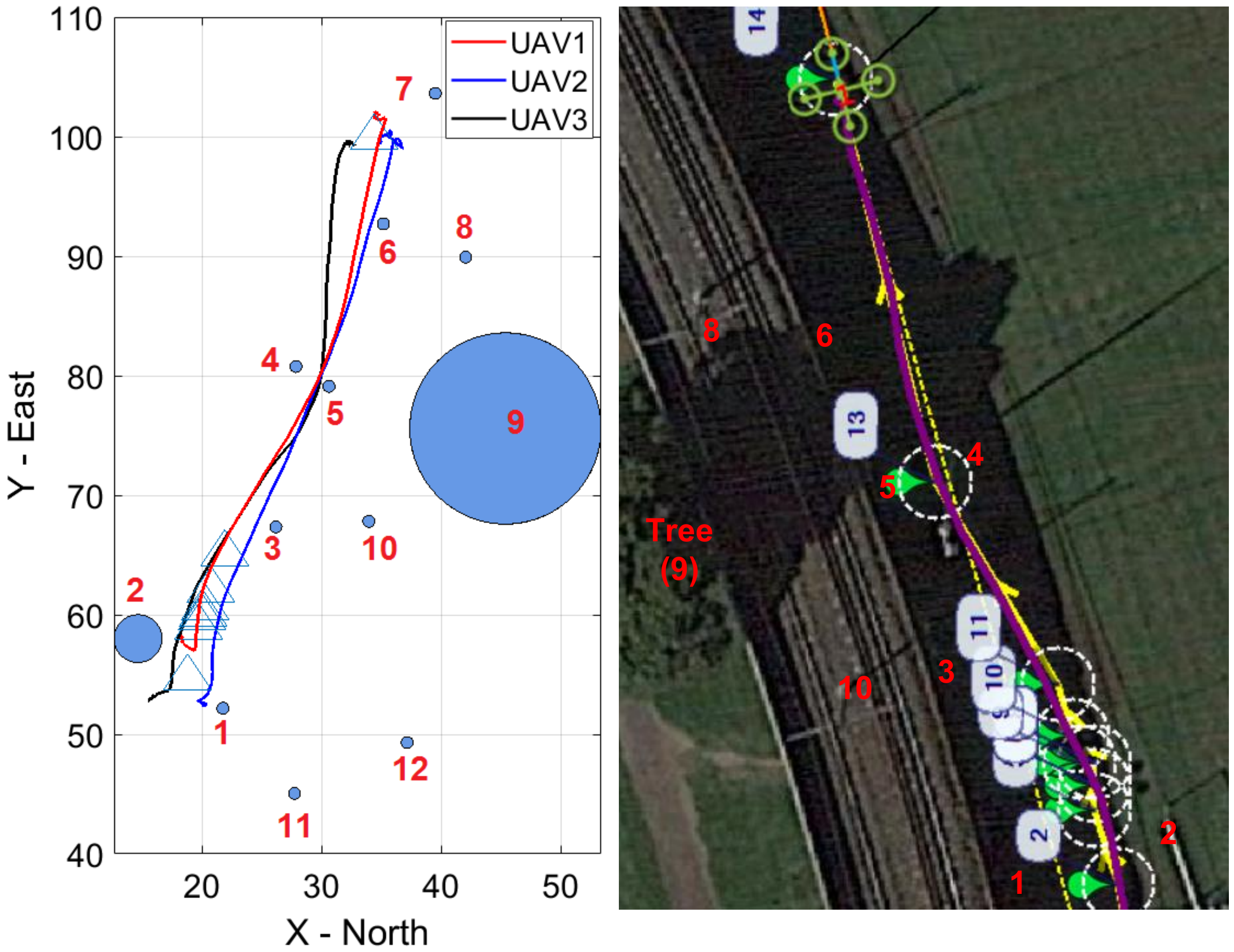}
	\caption{UAVs' trajectories in horizontal plane in the alignment formation}
	\label{exp_trans_align_xy}
	\vspace{-3mm}
\end{figure}

The evaluation is conducted in three reconfiguration shapes where the formation needs to change its configuration to keep safe while fulfilling the inspection task. The reference points are chosen to coincide with the centroid of the triangle created by the three UAVs. In experiments, it is planned that the designed formation shape starts to reconfigure at waypoint 11 and fully transforms into the new shape at about 1 m before waypoint 13. The new shape would be preserved until 1 m after the waypoint 13 and then complete the reconfiguration process at waypoint 14, which is illustrated in the right image of Fig. \ref{exp_trans_align_xy}.  Figure \ref{real_form} shows pictures that were captured from the field test of the formation transformation from the original horizontal triangle shape (Fig. \ref{real_form}.a) to the alignment (Fig. \ref{real_form}.b) and the vertical triangle (Fig. \ref{real_form}.c) ones. 

\begin{figure}[!htbp]
	\centering
	\includegraphics[width=8.5cm]{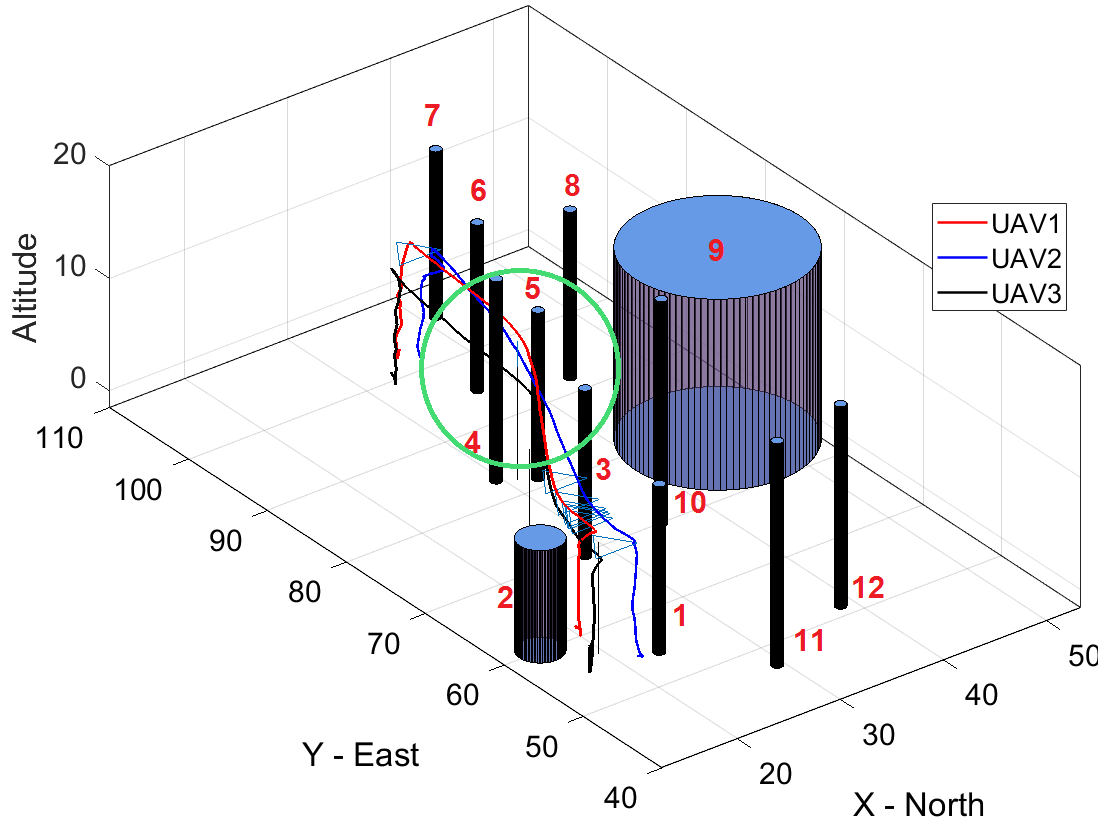}
	\caption{3D real-time plot for alignment formation}
	\label{exp_trans_align_3D}
	\vspace{-3mm}
\end{figure}

In the experiment with the alignment reconfiguration, Fig. \ref{exp_trans_align_3D} shows the capability of the formation in traversing the narrow corridor in which space is just enough for a single UAV to pass through. It shows clearly in the figure that the reconfiguration is completed to allow the UAVs to go through the narrow passage between obstacles 4 and 5 without any contact.

\begin{figure}[!htbp]
	\centering
	\includegraphics[width=8.5cm]{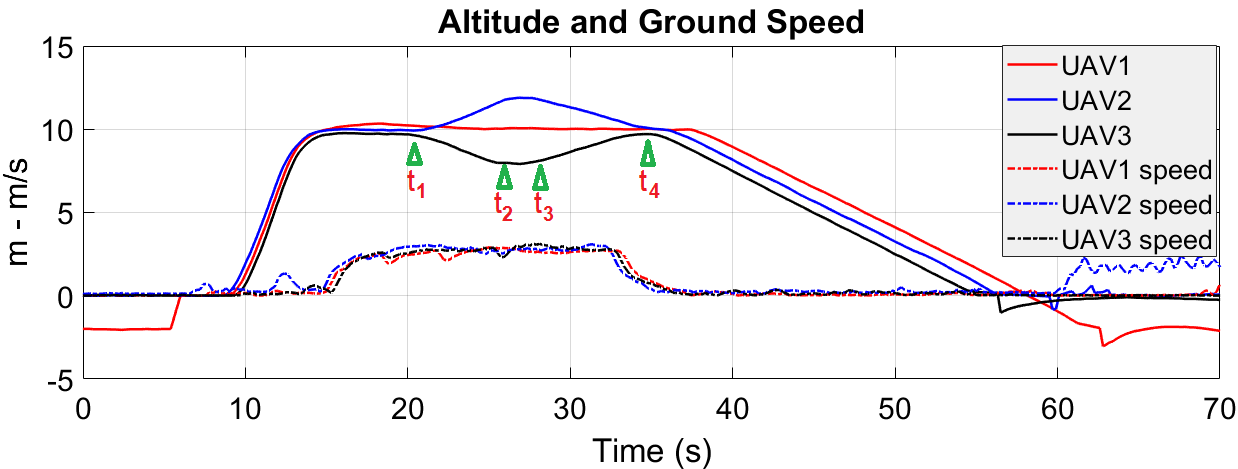}
	\caption{Altitudes and ground speeds of UAVs during the experiment with rotating reconfiguration}
	\label{Rot_AltVelo}
\end{figure}

\begin{figure}[!htbp]
	\centering
	\includegraphics[width=8.5cm]{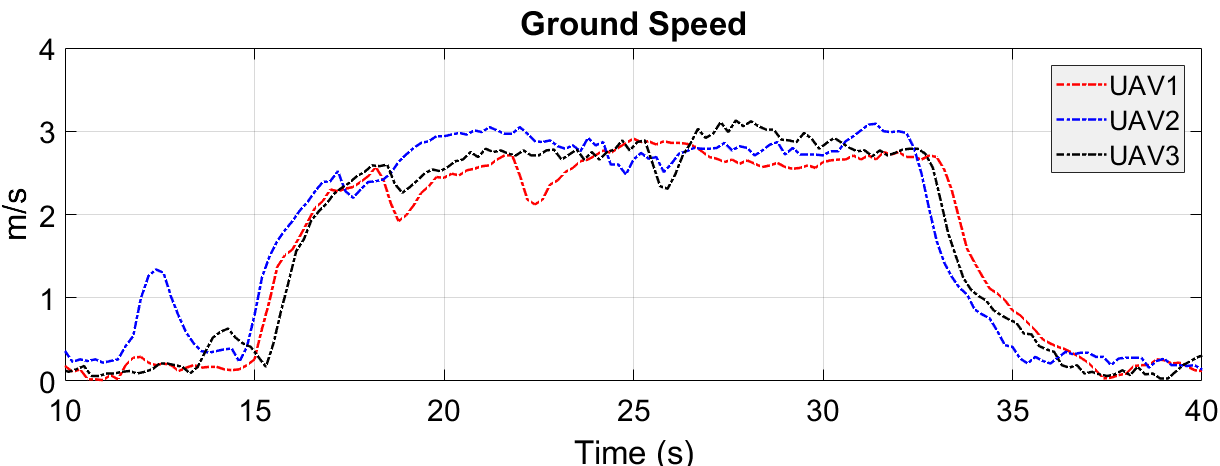}
	\caption{Ground speed of UAVs in rotating reconfiguration}
	\label{Rot_Velo}
\end{figure}

In the rotation transformation, the UAV$_1$ kept following its planned path while the two others changed their flight heights to reach their new positions in the vertical plane as shown in Fig. \ref{Rot_AltVelo}. Specifically, changes in the altitude happened at time $t_1 = 20$ s to reconfigure the UAVs fully to the new shape at time $t_2 = 26$ s. The new shape is then preserved until $t_3 = 27$ s and finally converted back to its original at $t_4 = 35$ s. The velocity profiles of the UAVs shown in Fig. \ref{Rot_Velo} imply the relatively stable movement of UAVs during this experiment.


\begin{figure}[!htbp]
	\centering
	\includegraphics[width=8.5cm]{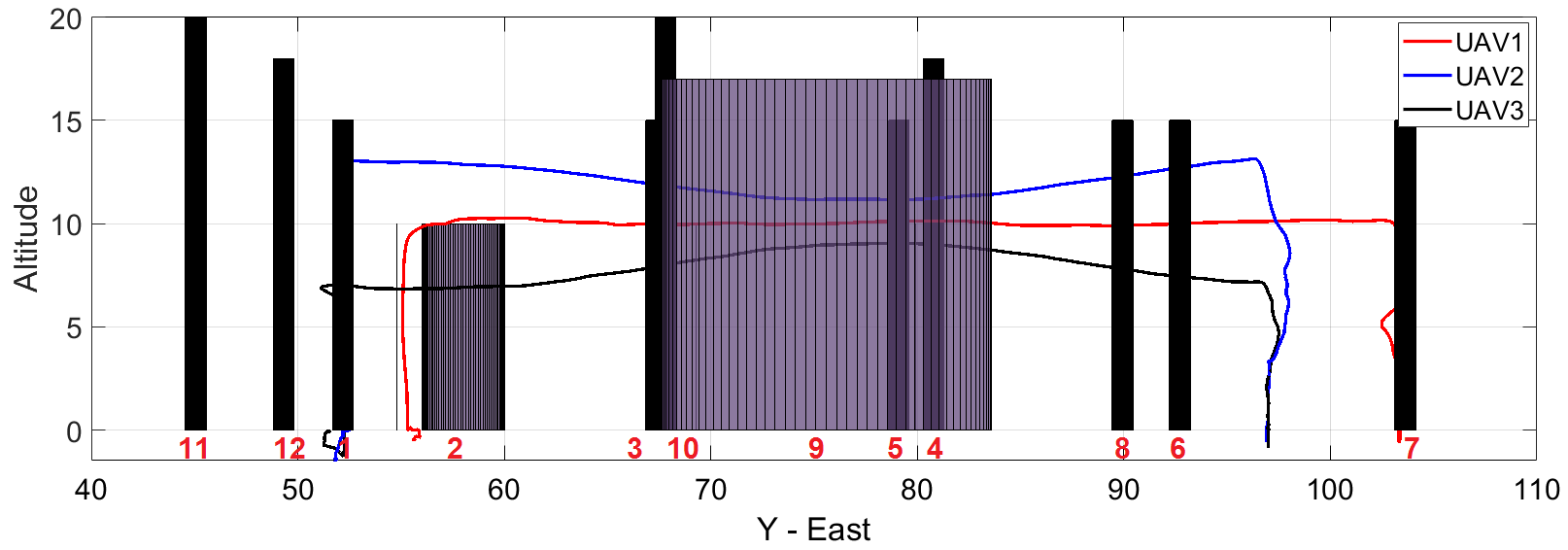}
	\caption{Trajectories of UAVs in the experiment with shrink reconfiguration}
	\label{trans_sim}
\end{figure}

Fig. \ref{trans_sim} shows the result of the shrink reconfiguration. The UAVs start to change their altitudes when encountering obstacle 3 and shrink the triangular formation to pass through obstacle 9. This result proves the capability of the proposed algorithm in handling situations where the formation shape needs to be maintained, but its size need adjusting to avoid collisions.


 On the other hand, it is also noted that the reconfiguration in experiments was conducted by using offline satellite maps. While this approach is relevant for most static civil infrastructure, occasionally unexpected dynamic obstacles not included in the calculation may cause safety concerns. The problem can be overcome by incorporating real-time data acquired by sensors installed on UAVs which will be our next focus. 

\section{CONCLUSIONS} \label{conclusion}

In this paper, we have proposed a path planning algorithm for multi-UAV formation in which its shape can vary in accordance with the operating environment. The core of our algorithm is the derivation of a cost function that takes into account the constraints on collision avoidance, flight altitude, communication range, and visual inspection requirements. Based on it, the $\theta$-PSO has been used to generate the path for the formation which is then used to determine trajectories for individual UAVs. We have also proposed the use of intermediate waypoints for reconfiguration which can be accomplished in the alignment, rotation, or shrink fashion. A number of experiments have been completed to evaluate the performance of the proposed algorithm for inspection tasks. The results obtained not only confirm its validity but also practically suggest a possibility of extending the work toward a generic reconfigurable architecture for robotic formation control in complex environments. 






\bibliography{bibi}

\begin{thebibliography}{10}
\providecommand{\url}[1]{#1}
\csname url@rmstyle\endcsname
\providecommand{\newblock}{\relax}
\providecommand{\bibinfo}[2]{#2}
\providecommand\BIBentrySTDinterwordspacing{\spaceskip=0pt\relax}
\providecommand\BIBentryALTinterwordstretchfactor{4}
\providecommand\BIBentryALTinterwordspacing{\spaceskip=\fontdimen2\font plus
\BIBentryALTinterwordstretchfactor\fontdimen3\font minus
  \fontdimen4\font\relax}
\providecommand\BIBforeignlanguage[2]{{%
\expandafter\ifx\csname l@#1\endcsname\relax
\typeout{** WARNING: IEEEtran.bst: No hyphenation pattern has been}%
\typeout{** loaded for the language `#1'. Using the pattern for}%
\typeout{** the default language instead.}%
\else
\language=\csname l@#1\endcsname
\fi
#2}}

\bibitem{rathinam2008vision}
S.~Rathinam, Z.~W. Kim, and R.~Sengupta, ``Vision-based monitoring of locally
  linear structures using an unmanned aerial vehicle,'' \emph{Journal of
  Infrastructure Systems}, vol.~14, no.~1, pp. 52--63, 2008.

\bibitem{Yan2017}
M.-d. Yan, X.~Zhu, X.-x. Zhang, and Y.-h. Qu, ``Consensus-based
  three-dimensional multi-uav formation control strategy with high precision,''
  \emph{Frontiers of Information Technology {\&} Electronic Engineering},
  vol.~18, no.~7, pp. 968--977, Jul 2017.

\bibitem{dong2015}
X.~Dong, B.~Yu, Z.~Shi, and Y.~Zhong, ``Time-varying formation control for
  unmanned aerial vehicles: Theories and applications,'' \emph{IEEE
  Transactions on Control Systems Technology}, vol.~23, no.~1, pp. 340--348,
  2015.

\bibitem{he2018}
L.~He, P.~Bai, X.~Liang, J.~Zhang, and W.~Wang, ``Feedback formation control of
  uav swarm with multiple implicit leaders,'' \emph{Aerospace Science and
  Technology}, vol.~72, pp. 327--334, 2018.

\bibitem{kim2006}
D.~H. Kim, H.~Wang, and S.~Shin, ``Decentralized control of autonomous swarm
  systems using artificial potential functions: Analytical design guidelines,''
  \emph{Journal of Intelligent and Robotic Systems}, vol.~45, no.~4, pp.
  369--394, 2006.

\bibitem{do2007}
K.~D. Do, ``Bounded controllers for formation stabilization of mobile agents
  with limited sensing ranges,'' \emph{IEEE Transactions on Automatic Control},
  vol.~52, no.~3, pp. 569--576, 2007.

\bibitem{Liao2017}
F.~{Liao}, R.~{Teo}, J.~L. {Wang}, X.~{Dong}, F.~{Lin}, and K.~{Peng},
  ``Distributed formation and reconfiguration control of vtol uavs,''
  \emph{IEEE Transactions on Control Systems Technology}, vol.~25, no.~1, pp.
  270--277, Jan 2017.

\bibitem{semsar2008}
E.~Semsar-Kazerooni and K.~Khorasani, ``Optimal consensus algorithms for
  cooperative team of agents subject to partial information,''
  \emph{Automatica}, vol.~44, no.~11, pp. 2766--2777, 2008.

\bibitem{panagou2014}
D.~Panagou and V.~Kumar, ``Cooperative visibility maintenance for
  leader--follower formations in obstacle environments,'' \emph{IEEE
  Transactions on Robotics}, vol.~30, no.~4, pp. 831--844, 2014.

\bibitem{yu2016}
X.~Yu and L.~Liu, ``Distributed formation control of nonholonomic vehicles
  subject to velocity constraints,'' \emph{IEEE Transactions on Industrial
  Electronics}, vol.~63, no.~2, pp. 1289--1298, 2016.

\bibitem{Nguyen2004}
A.~D. Nguyen, Q.~P. Ha, S.~Huang, and H.~Trinh, ``Observer-based decentralised
  approach to robotic formation control,'' in \emph{Proc. Australian Conf.
  Robotics and Automation}, 2004, pp. 1--8.

\bibitem{sharma2009}
R.~Sharma and D.~Ghose, ``Collision avoidance between uav clusters using swarm
  intelligence techniques,'' \emph{International Journal of Systems Science},
  vol.~40, no.~5, pp. 521--538, 2009.

\bibitem{wang2013}
J.~Wang and M.~Xin, ``Integrated optimal formation control of multiple unmanned
  aerial vehicles,'' \emph{IEEE Transactions on Control Systems Technology},
  vol.~21, no.~5, pp. 1731--1744, 2013.

\bibitem{wang1998}
P.~Wang and F.~Hadaegh, ``Optimal formation-reconfiguration for multiple
  spacecraft,'' in \emph{Guidance, Navigation, and Control Conference and
  Exhibit}, 1998, p. 4226.

\bibitem{wang1999}
K.~Wang, ``Minimum-fuel formation reconfiguration of multiple free-flying
  spacecraft,'' \emph{J. of Astronautical Sciences}, vol.~47, no.~1, pp.
  77--102, 1999.

\bibitem{sauter2012}
L.~Sauter and P.~Palmer, ``Onboard semianalytic approach to collision-free
  formation reconfiguration,'' \emph{IEEE Transactions on Aerospace and
  Electronic Systems}, vol.~48, no.~3, pp. 2638--2652, 2012.

\bibitem{duan2010}
H.~Duan and S.~Liu, ``Non-linear dual-mode receding horizon control for
  multiple unmanned air vehicles formation flight based on chaotic particle
  swarm optimisation,'' \emph{IET control theory \& applications}, vol.~4,
  no.~11, pp. 2565--2578, 2010.

\bibitem{sui2017}
Z.~Sui, Z.~Pu, and J.~Yi, ``Optimal uavs formation transformation strategy
  based on task assignment and particle swarm optimization,'' in \emph{2017
  IEEE International Conference on Mechatronics and Automation (ICMA)}.\hskip
  1em plus 0.5em minus 0.4em\relax IEEE, 2017, pp. 1804--1809.

\bibitem{duan2013}
H.~Duan, Q.~Luo, Y.~Shi, and G.~Ma, ``Hybrid particle swarm optimization and
  genetic algorithm for multi-uav formation reconfiguration,'' \emph{IEEE
  Computational intelligence magazine}, vol.~8, no.~3, pp. 16--27, 2013.

\bibitem{ma2010}
G.~Ma, H.~Huang, and Y.~Zhuang, ``Time optimal trajectory planning for
  reconfiguration of satellite formation with collision avoidance,'' in
  \emph{Control and Automation (ICCA), 2010 8th IEEE International Conference
  on}.\hskip 1em plus 0.5em minus 0.4em\relax IEEE, 2010, pp. 476--479.

\bibitem{wang2012}
S.~Wang and C.~Zheng, ``A hierarchical evolutionary trajectory planner for
  spacecraft formation reconfiguration,'' \emph{IEEE Transactions on Aerospace
  and Electronic Systems}, vol.~48, no.~1, pp. 279--289, 2012.

\bibitem{paul2008}
T.~Paul, T.~R. Krogstad, and J.~T. Gravdahl, ``Uav formation flight using 3d
  potential field,'' in \emph{Control and Automation, 2008 16th Mediterranean
  Conference on}.\hskip 1em plus 0.5em minus 0.4em\relax IEEE, 2008, pp.
  1240--1245.

\bibitem{feng2018}
Y.~Feng, Y.~Wu, H.~Cao, and J.~Sun, ``Uav formation and obstacle avoidance
  based on improved apf,'' in \emph{2018 10th International Conference on
  Modelling, Identification and Control (ICMIC)}.\hskip 1em plus 0.5em minus
  0.4em\relax IEEE, 2018, pp. 1--6.

\bibitem{seo2012}
J.~Seo, Y.~Kim, S.~Kim, and A.~Tsourdos, ``Consensus-based reconfigurable
  controller design for unmanned aerial vehicle formation flight,''
  \emph{Proceedings of the Institution of Mechanical Engineers, Part G: Journal
  of Aerospace Engineering}, vol. 226, no.~7, pp. 817--829, 2012.

\bibitem{jian2016}
W.~Jian-hong and R.~Javed~Masood, ``Interior point algorithm for multi-uavs
  formation autonomous reconfiguration,'' \emph{Journal of Control Science and
  Engineering}, vol. 2016, 2016.

\bibitem{hoang2018iros}
V.~T. Hoang, M.~D. Phung, H.~D. Tran, and Q.~P. Ha, ``Angle-encoded swarm
  optimization for uav formation path planning,'' in \emph{Intelligent Robots
  and Systems (IROS), 2018 IEEE/RSJ International Conference on}.\hskip 1em
  plus 0.5em minus 0.4em\relax IEEE, 2018, pp. 5239--5244.

\bibitem{Hoang2017AT}
V.~Hoang, M.~D. Phung, and Q.~Ha, ``Adaptive twisting sliding mode control for
  quadrotor unmanned aerial vehicles,'' in \emph{2017 11th Asian Control
  Conference (ASCC)}.\hskip 1em plus 0.5em minus 0.4em\relax IEEE, 2017, pp.
  671--676.

\end{thebibliography}

\end{document}